*Research Article*

# Early Diagnosis of Retinal Blood Vessel Damage via Deep Learning-Powered Collective Intelligence Models

**Pranjal Bhardwaj,**[1] **Prajjwal Gupta**[iD],[1] **Thejineaswar Guhan**[iD],[2] **and Kathiravan Srinivasan**[iD][1]

[1]*School of Computer Science and Engineering, Vellore Institute of Technology, Vellore 632 014, India*
[2]*School of Information Technology and Engineering, Vellore Institute of Technology, Vellore 632 014, India*

Correspondence should be addressed to Kathiravan Srinivasan; kathiravan.srinivasan@vit.ac.in

Early diagnosis of retinal diseases such as diabetic retinopathy has had the attention of many researchers. Deep learning through the introduction of convolutional neural networks has become a prominent solution for image-related tasks such as classification and segmentation. Most tasks in image classification are handled by deep CNNs pretrained and evaluated on imagenet dataset. However, these models do not always translate to the best result on other datasets. Devising a neural network manually from scratch based on heuristics may not lead to an optimal model as there are numerous hyperparameters in play. In this paper, we use two nature-inspired swarm algorithms: particle swarm optimization (PSO) and ant colony optimization (ACO) to obtain TDCN models to perform classification of fundus images into severity classes. The power of swarm algorithms is used to search for various combinations of convolutional, pooling, and normalization layers to provide the best model for the task. It is observed that TDCN-PSO outperforms imagenet models and existing literature, while TDCN-ACO achieves faster architecture search. The best TDCN model achieves an accuracy of 90.3%, AUC ROC of 0.956, and a Cohen's kappa score of 0.967. The results were compared with the previous studies to show that the proposed TDCN models exhibit superior performance.

## 1. Introduction

Diabetic retinopathy (DR) is a medical condition caused due to complications caused by diabetes mellitus that influences the ocular perceivers and causes damage to the delicate tissues of the retina. This condition can occur in any adult who is suffering from either type 1 or type 2 diabetes. As diabetic retinopathy increases with time, it eventually causes a complete loss of vision. In this paper, we will be discussing the four different types of DR (nonproliferative DR, maculopathy, preproliferative, and proliferative). With the rise in cases for DR, there arises a need for automation for the detection of diabetic retinopathy in fundus images. Identification of individual characteristics and extraction of features are important for the assessment of eye disorders such as DR and other retinal diseases.

Manual inspection of fundus images can prove to be a tedious process to decipher subtle variations in microaneurysms, optic disks, hemorrhages, blood vessels, hard exudates, soft exudates, and macular edema. In such situations, CAD (computer-aided diagnostic) systems can significantly reduce the manual inspection load for professionals. These methods also reduce margins of error compared to ophthalmologists while examining the fundus images. With the introduction of convolutional neural networks, deep learning has become the de facto for computer vision-based problems. CNNs have the capability to extract features automatically and decipher patterns for better understanding of the data. Deep CNN models are large deep learning models which leverage the power of various CNNs and pooling layers. These models typically stack a bunch of CNNs together to form a feed-forward network and often produce state-of-the-art results. Most of the deep CNN models are trained and tested on the imagenet classification dataset, and the versatility of deep learning enables it to transfer learning from one dataset to another. This aspect of transfer learning has made most computer vision problems efficient and solvable through training imagenet models on a dataset [1–5]. The imagenet models are usually employed with approaches like transfer learning.

Although transfer learning is a powerful concept, models trained on imagenet often do not obtain the best result when trained on an independent dataset. The reasons for the same could include the size of the dataset, the size of the model, and the type of the data, thus bringing the need for tailor-made models. Models can be proposed on the basis of the dataset, where a small dataset may require a lightweight model. With the help of lightweight models, better performance and high scalability can be achieved. In such a situation, tailor-made models could possibly be the most efficient model for the job. However, building tailored models can be computationally expensive. Moreover, in certain cases, the features extracted from the imagenet dataset would not be a great starting point, the reason for the same being the dissimilarity of the images in the imagenet dataset and the dataset on which the model is to be trained. For example, the imagenet dataset contains objects such as cars and animals which do not resemble fundus images.

Nature-inspired swarm algorithms have been heavily employed for feature selection on various tasks. These algorithms have performed efficiently in hyperparameter search/optimization. Hyperparameter optimization methods such as grid search have been computationally expensive methods and give the best result within a search space. Hence, swarm algorithms have been used for architectural proposals. Nature-inspired swarm algorithms are a subset of evolutionary algorithms. These algorithms in general have been known to shrink the search space whether it be feature selection or hyperparameter optimization. However, the usage of swarm-based algorithms has been limited to feature selection as demonstrated across in [6–8]. Most used cases of swarm-related algorithms in DR classification are restricted to either hyperparameter optimization or feature extraction. This methodology utilizes both methods by proposing new models with various hyperparameters based on the dataset. TDCNs are the outcomes of the swarm algorithms, and in this case, 2 algorithms are used. The objective of these models is to provide custom models which are efficient and accurate for a given dataset. The proposed models rely solely on architecture rather than fixed models with a set starting weight to achieve accuracy. Therefore, the significance of transfer learning diminishes which addresses the problems caused by imagenet models.

The major contributions of this paper include the following:

(i) A formal definition of search space complexity for searching tailor-made models

(ii) Utilizing two swarm intelligence algorithms, namely, ant colony optimization (ACO) and particle swarm optimization (PSO), to perform heuristic architecture search in multidimensional space

(iii) Two tailored ConvNets (TDCN) called TDCN-ACO and TDCN-PSO were obtained from ACO- and PSO-based searches. TDCN-PSO being significantly smaller (lightweight) achieves better performance over 3 metrics, accuracy, AUC ROC, and Cohen's kappa, than the imagenet models. The model's performance is solely defined on the basis of architecture search (a dynamic process) rather than designing a fixed model

(iv) TDCN-PSO and TDCN-ACO produce competitive results with many proposed architectures in literature for fundus image classification

## 2. Related Works

Over the years, various deep learning approaches have been utilized to diagnose retinal vascular diseases, including usage of imagenet models, transfer learning, and image segmentation. Kanungo et al. [9] proposed an Inception-V3 architecture-based approach in which multiple convolution filter inputs are processed on the same input. This takes advantage of multilevel feature extraction from each input. The accuracy achieved is comparatively less than the other approaches based on similar architectures. Zeng et al. [10] proposed a deep Siamese CNN architecture for automatic diabetic retinopathy classification. The model takes two fundus images as input, i.e., the left and right eye image. Transfer learning approaches have shown great results in terms of DR classification. Lam et al. [11] proposed an approach using GoogleNet and AlexNet to demonstrate transfer learning. Sakaguchi et al. [12] used GNNs to construct the region of interest graphs which is helpful for further classification of DR using CNN. Although imagenet and transfer learning-based approaches perform well on benchmark datasets, they do not perform similarly on some niche datasets as they were designed keeping a general purpose in mind. In order to increase the performance of the model, the need for tailoring models arises.

Hyperparameter optimization (HO) plays a key role when designing tailor-made models for specific datasets. Over the years, automated hyperparameter search algorithms and libraries have become increasingly popular; the widely used ones include hyperopt [13], optuna [14], and Auto WEKA [15]. The total search space including all subsets is $2^n$ where $n$ is the number of features, thus posing a high-dimensional problem. The early works focused on using Bayesian optimization for hyperparameter tuning [16–18], but these techniques suffer the drawback of being computationally complex. Li et al. address the problem of HO as an exploration-based nonstochastic infinite-armed bandit problem and proposed hyperband to solve the same [19]. Falkner et al. addressed the robustness and scalability of HO [20], and Hazan et al. addressed the high-dimensional problems [21], each work resulting in a manifold increase in performance over hyperband. Hintz et al. pointed out that a significant speedup can be achieved by carrying out the search on lower-dimensional data representation at the beginning and increasing the dimensions later in the optimization process [22]. Some of the recent works revolve around using reinforcement learning for adaptive and dynamic HO in contrast to other approaches which require the user to set generic ranges in order to define the HO search space [23, 24].

Nature-inspired swarm-based techniques are a subset of swarm intelligence which in turn is a subset of metaheuristic algorithm. Nature-inspired swarm algorithms have been of great use when it comes to feature selection in the tabular dataset as it achieves optimal results with substantially lower computational resources [25]. This behavior of swarm-based techniques has been taken to advantage in hyperparameter selection which is a computationally expensive process as mentioned before. Bee colony optimization and dragonfly algorithm were used by Yasen and Al-Madi [26] to select the size of hidden nodes in the neural network. Crossover and mutation were also combined to devise new models. All in all, the models were trained on medical patients' data, and the comparative results were obtained. Similarly, Sun et al. [27] used PSO to determine the dimensionality of the autoencoder to be used, moving to the applications of swarm-based algorithms on the application of retinal fundus images.

Bajčeta et al. [7] proposed a segmentation model of blood vessels where they have applied ACO in fundus images, and ACO performs feature extraction. Hooshyar and Khayati [8] proposed a mathematical model in which they have used eigenvalues of the Hessian matrix. They used a Gabor filter bank to extract the features from retinal images. For classification purposes, fuzzy c-means and ACO models are produced. Asad et al. [28] proposed an approach to ensure improved ant clustering-based segmentation. The approach was developed with the help of a new heuristic function of the ACO algorithm. Kavitha and Ramakrishnan [29] employed an ACO algorithm using an OTSU method which is based on the inputs from optic disc and macula. Balakrishnan et al. [6] came up with a hybrid model for classification and feature extraction from retinal images. Channel extraction and median filter are used for preprocessing. After the preprocessing stage, the authors used a histogram of oriented gradient (HOG) with a complete local binary pattern (CLBP) to perform feature selection. Particle swarm optimization along with fuzzy membership functions is used by Bhimavarapu et al. [30] to cluster images, and then, probabilistic models were used to segment fundus images. Although nature-inspired swarm-based methods have been extensively studied on hyperparameter optimization, its application is restricted to feature selection alone in diabetic retinopathy. Furthermore, identifying intrinsic features in medical datasets has led researchers to build complex deep architectures which are computationally expensive. For faster convergence and inference, there is a requirement of lightweight models which can be achieved using the likes of nature-inspired swarm algorithms as discussed in further sections.

## 3. Methodology

The process of generation of a tailored model can be defined as a search on a set of models $S_m$ formed by the combinations of layers from a set of layers $S_l$ using a selection metric $f$ and a goal $G$. For a given $S_m$, the search is deemed successful if the set of resultant models $S_r$ is a subset of the set of models $S_r$ which satisfy $G$ on $f$ with an acceptable performance bound $\epsilon$.

$$S_r = \{m : m \in S_m, f(m) + \epsilon \vdash G\}. \quad (1)$$

The complexity of the search is determined by the cardinality of the set $S_m(t_s)$. For a given bound of the minimum number of layers $B_l$ and a maximum number of layers $B_u$, $t_s$ can be obtained by

$$t_s = \sum_{b=B_l}^{B_u} t_l^b, \quad (2)$$

where $t_l$ is the cardinality of the set $s_l$.

This search space can be restricted by using various design parameters (order, count, blocks, etc.) on the occurrence of layers. This search space evidently becomes very large as $t_l$ and $B_l$ increase, and simple techniques like a brute force search become time and resource intensive. This problem can be solved using heuristic-based search algorithms, such as swarm intelligence algorithms as demonstrated further in this section.

For the two swarm-based techniques discussed in this paper (ACO and PSO), we obtain a single model ($S_s$) as the result of the search. We tested the results on three selection metrics as discussed in Section 4. The search goal $G$ was set to the best result obtained in the case of imagenet models or better essentially resulting in a search which yields lightweight models of similar performance to imagenet models or high performance models or both. The value of $\epsilon$ was set such that the search resulted in a model which performed better than at least one of the imagenet models. The nature of results is characterized by the settings of $B_l$ and $B_u$. The detailed success criterion is listed in Table 1.

The resultant architectures, TDCN-ACO, and TDCN-PSO have been described in the following sections using the abbreviations in Table 2.

*3.1. TDCN-ACO Search Using Ant Colony Optimization.* Ants have inspired many methods and techniques, among which the most studied is the general purpose optimization technique known as ACO (ant colony optimization). ACO is inspired by the behavior of some ant species. These ants deposit pheromone on the ground to mark a favorable path for the rest of the members of the colony. Each ant moves randomly and more pheromone gets deposited on the path. The more the pheromone on the path, the higher is the probability of the path being followed. In this work, the implementation proposed by Byla and Pang [31] is used.

The search for models is based on the number of maximum ants and maximum search depth; the algorithm starts the process by creating an internal graph with just an input node. Sequentially, ants are created until the number of maximum ants allowed is created. These ants follow ant colony selection (ACS) rule, wherein the ants choose a particular layer based on the algorithmic calculation in ACO. If the chosen layer has not been chosen by a previous ant, then the layer is added as a neighbor node to the node chosen by the previous ant. Based on the same rule, the parameters of the layer are also chosen from a specified list of parameters. Once the choice is made, ACS local pheromone update is done by the ant. The same is done until the maximum number of ants allowed, traverses through to the same depth. The paths are converted to neural networks, and based on the evaluation metric, the best ant is chosen, and now, an ACS global pheromone update is done. Note that the previous update is kept local to encourage the ants to pave new paths. Figure 1 conveys the same in a pictorial fashion.

For this dataset, 4 hyperparameters were mainly handled:

(i) *Number of ants*: 8 and 16. Based on the experiments conducted by [31], it was found that the greatest decrease in error was identified when the number of ants was increased from 8 to 16. 32 may yield better results, but the duration of computation will increase

(ii) *Search depth*: 32. Most of the available image models tend to be deep models; therefore, a search depth of 32 was used

(iii) *Epochs*: it is kept at 10 epochs

(iv) *Image size*: 32 and 64. When large image sizes are used, the accuracy more or less remains the same. This was surprising considering the performance output with respect to the image size

The permutations among the defined hyperparameters have generated 4 models. The trend across the models happens to look like this, C2D | BN till one max pooling, and lastly followed by a series of DE | BN layers. The choice of layers across models is almost comparable to the traditional model building approach, where usually every CN2D layer is succeeded by a pooling such as avg. pooling. However, the trend of using a pooling layer before flattening the outputs of CNN has generally shown success, and this process utilizes that.

As far as the search space is concerned, a total of 4 layers were permuted across C2D, MP or AP, BN, and DE. The search space for every layer varies as per requirements:

(i) C2D: 9 permutations possible

(ii) MP: 2 permutations possible

(iii) BN: 1 permutation possible

(iv) DE: 4 permutations possible

The number of permutations can be maintained within a permutation matrix which can be accessed simultaneously when a layer is accessed $L = [C2D, MP, BN, DE]$ and $= [9, 2, 1, 4]$. So if a layer $L_i$ is chosen, then $P_i$ is also chosen, where $i$ is the index in the array $L$. So the search space when $n$ is the number of layers is shown in

$$\text{Search space} = t_s = P_0^{\text{num C2D}} + P_1^{\text{num MP}} + P_2^{\text{num BN}} + P_3^{\text{num DE}}. \quad (3)$$

Table 1: Search goals and performance bounds used in this work.

| Metric | $G$ | $\epsilon$ |
| --- | --- | --- |
| Accuracy | ≥74.8 | 1.5% |
| AUC ROC | ≥0.91 | 0.2 |
| Cohen's kappa | ≥0.776 | 0.02 |

The search space is model independent and considering a model of 12 C2D | BN then MP and lastly 3 DE | BN layers gives a search space as shown in

$$t_s = 282,429,569,261. \quad (4)$$

In total, there exist 16 ants at maximum denoted by *ant* for a total *depth* of 32. Equation (5) conveys the number of models which were trained in the tailor fitting process:

$$\text{Number of models trained} = \text{ants} * \text{depth} = 16 * 32 = 512. \quad (5)$$

*3.2. TDCN-PSO Search Using Particle Swarm Optimization.* Particle swarm optimization is a nature-inspired, population-based algorithm that has been found to optimize or converge better than standard optimization algorithms like conjugate descent, gradient descent, and Newton method in multidimensional space. Stochastic search is used to achieve the same by moving the particles (candidate solutions) in the search space using mathematical computations to calculate the velocity and direction of the particle swarm by considering the particle's current position and goal. A global best (in terms of position) is maintained as well as a local best for each particle is maintained which influences in making decisions for the particle's movement in consequent iterations. This results in the general behavior of the swarm moving towards the best solutions for the given search space.

The position and velocity after each generation or iteration of the algorithm can be obtained by

$$X^{t+1}ij = X^t ij + V^{t+1}_{ij}, \quad (6)$$

$$V^{t+1}ij = wV^t ij + c_1 r^t 1 \left(\text{pbest}ij - X^t ij\right) + c_1 r^t 2 \left(\text{gbest}j - X^t_{ij}\right), \quad (7)$$

where $X^t_i$ is the position vector of $i^{\text{th}}$ particle at $i^{\text{th}}$ and $V^t_i$ is the position vector of $i^{\text{th}}$ particle at $i^{\text{th}}$.

Many modifications to these equations have been proposed over time which offers specific benefits of their own, but the original algorithm is proceeded across for a general comparison of this algorithm with others.

Table 2: ConvNet layer abbreviations used in this work.

| Abbreviation | Layer |
|---|---|
| C2D | Conv2D (convolutional layer with 2-dimensional filters) |
| BN | Batch normalization |
| DO | Dropout |
| MP | Max pooling |
| AP | Average pooling |
| DE | Dense or fully connected |
| F | Flatten |

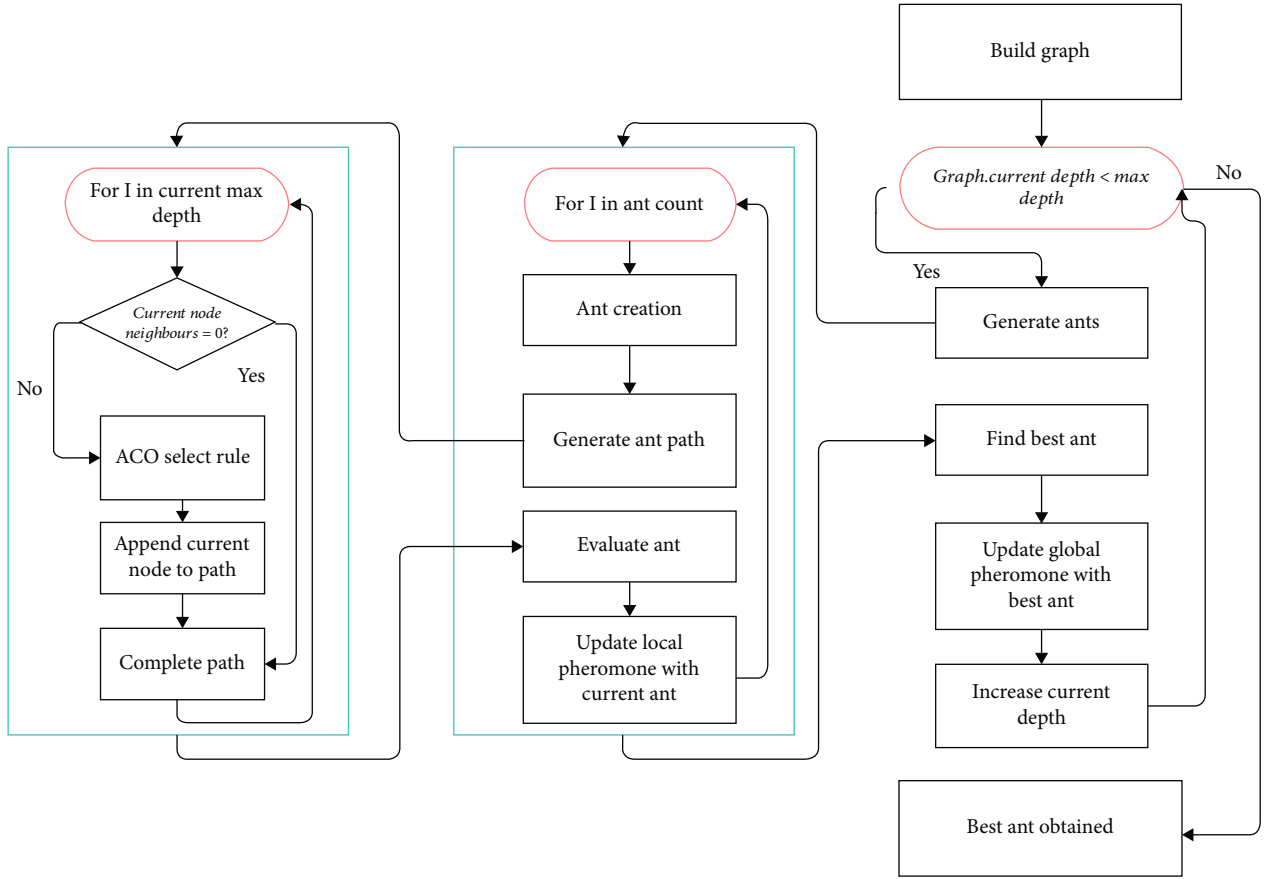

Figure 1: TDCN-ACO search using ant colony optimization.

The tailoring of models is carried out by taking several parameters into consideration such as population size, number of iterations and runs, the minimum and maximum number of layers, and probability of each type of layer. Using these parameters, the particles are initialized by assigning them random CNN architectures represented in an array form that satisfy the conditions and constraints defined by the parameters (number of layers, layer probabilities, kernel sizes, and so on). Then, each particle is fit and evaluated to find the local and global best based on the accuracy obtained on the validation data. The good blocks in the global best are passed onto further generations, and each particle is evaluated again in every iteration. An architecture search using PSO mainly involves six procedures, namely, representation of particles in form of architectures (in our case deep ConvNets) of varying properties, initializing the particles, evaluating the fitness of individual particles, a measure of the difference between two particles, computation of velocity, and updating the particles as illustrated in Figure 2.

In this paper, the search on a set formed by the combination of 4 different kinds of layers is performed, namely, convolutional, max pooling, average pooling, and fully connected, making $t_l$ as 4. We set the bounds $B_l = 3$ and $B_u = 20$, effectively giving us equation (8).

$$t_s = \sum_{b=3}^{20} 4^b \Rightarrow t_s = 1{,}466{,}015{,}503{,}680. \qquad (8)$$

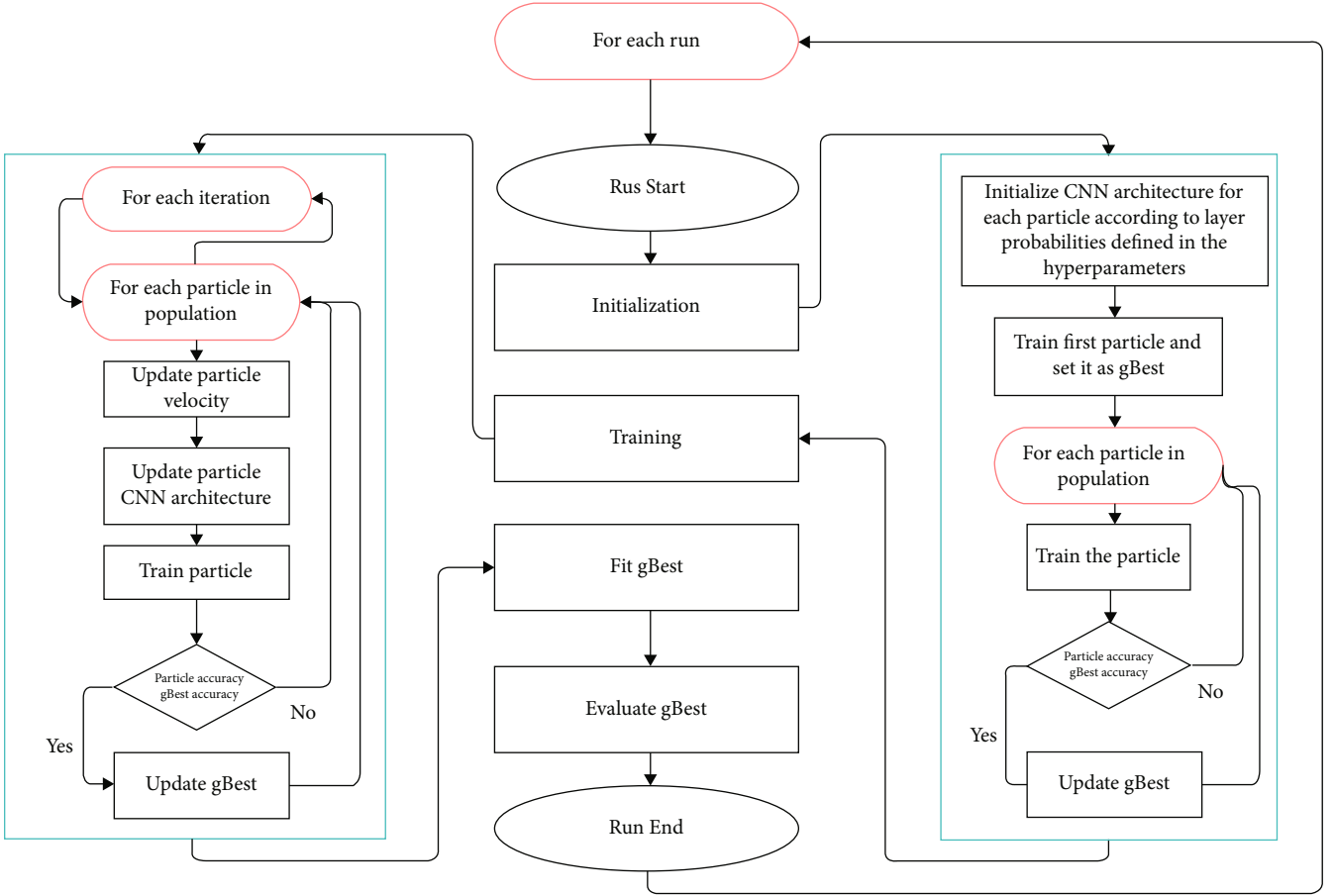

Figure 2: TDCN-PSO search using particle swarm optimization.

This means that the search space for the algorithm consists of more than 1466 billion models when no design constraints are applied. A simple design constraint was kept as follows:

(i) At least 1 DE layer

(ii) F has occurred before the first DE layer

(iii) No C2D layer after F layer

This decreased the search space manifolds. The number of models searched by the PSO algorithm ($S_{m,\text{pso}}$) is a function of the number of iterations $i$, number of runs, and swarm size $p$ and is given by

$$S_{m,pso} = r \times i \times p. \quad (9)$$

For this paper, the following hyperparameters are set $r = 5$, $i = 12$, and $p = 20$, which gives $S_{m,pso} = 1200$. These values depend on the quality of TDCN required hardware availability and time constraints.

Note that each updated particle was considered as a new model. To carry out the architecture search, the implementation proposed by Junior et al. [32] was used.

TDCN-PSO is the model represented by the global best particle at the end of the architecture search. The hyperparameters used for the search can be found in Table 3. The resultant architecture has some peculiar features such as 4 consecutive pooling layers and irregular batch normalization and dropout layers, all of which are characteristics of a model unique to the dataset. The architecture of TDCN-PSO found for APTOS dataset is as follows: C2D | C2D | BN | DO | C2D | BN | MP | AP | MP | MP | DO | C2D | BN | F | DO | DE | BN.

## 4. Experimental Setup

The dataset that we used for this paper is "ATPOS 2019 Blindness Detection" from a Kaggle competition conducted in 2019. This dataset contains high-resolution images of the retina taken in different lighting conditions, with left and right fields for every test subject. The images have been captured through varying devices, offering different scenarios and generalizations of the developed model for adoption by a wide variety of hardware. The images present in the data are of different orientations too, some are shown anatomically (macula on left, optic nerve on right for right eye), and others are shown inverted. The training set consists of 3662 images, while the test set contains 1928 images.

Table 3: Hyperparameter setting for TDCN-PSO search.

| Category | Hyperparameter | Value |
| --- | --- | --- |
| Particle swarm optimization | Number of runs ($r$) | 5 |
| | Number of iterations ($i$) | 12 |
| | Swarm size ($p$) | 20 |
| | Cg | 0.5 |
| CNN architecture initialization | Minimum number of outputs from a Conv layer | 3 |
| | Maximum number of outputs from a Conv layer | 256 |
| | Minimum number of neurons in a FC layer | 1 |
| | Maximum number of neurons in a FC layer | 300 |
| | Minimum size of a Conv kernel | $3 \times 3$ |
| | Maximum size of a Conv kernel | $7 \times 7$ |
| | Minimum number of layers $B_l$ | 3 |
| | Maximum number of layers $B_u$ | 20 |
| | Dropout rate | 0.5 |
| Training | No. of epochs for the global best | 100 |
| | No. of epochs for particle evaluation | 1 |
| | Bath normalize layer outputs | Yes |
| Probability | Probability of convolutional layer | 0.7 |
| | Probability of pooling layer | 0.15 |
| | Probability of fully connected layer | 0.15 |

Table 4: Classes in APTOS 2019 dataset.

| Class ID | Class name | Number of samples |
| --- | --- | --- |
| 0 | No DR | 1805 |
| 1 | Mild | 370 |
| 2 | Moderate | 999 |
| 3 | Severe | 193 |
| 4 | Proliferative DR | 295 |

Table 5: Comparison of TDCN models with imagenet models and literature.

| Model | Accuracy | AUC ROC | Cohen's kappa |
| --- | --- | --- | --- |
| Inception | 73.2 | 0.91 | 0.738 |
| Xception | 74.8 | 0.87 | 0.772 |
| Resnet50 | 73.8 | 0.89 | 0.776 |
| Shaban.et al. [33] | 88 | 0.930 | 0.910 |
| S. Kassani et al. [34] | 83.09 | 0.950 | 0.892 |
| Taufiqurrahman et al. [35] | 85 | 0.820 | 0.925 |
| Bodapati et al. [36] | 84.31 | 0.970 | 0.758 |
| TDCN-ACO | 78.4 | 0.907 | 0.795 |
| TDCN-PSO | 90.3 | 0.956 | 0.967 |

The images are categorized into 5 classes (Table 4). Thus, this dataset poses a classification problem. The preprocessing strategies applied involve resizing.

Resizing: resizing in the digital image is changing the horizontal and vertical resolution of an image. Since the input images are in different resolutions, coming from different hardware, we decided to resize the images to $256 \times 256$ for the imagenet models, $32 \times 32$ and $64 \times 64$ for ACO, and $128 \times 128$ for PSO-CNN. Smaller image size was used for the swarm algorithms because of the high computational complexity and the time required to train the models.

The dataset is split into 10-fold cross validation. The following method is used to preserve the class wise distribution while evaluating the model. Also, this gives considerate weightage to classes, such as severe and proliferate DR, with lower number of samples.

To compare the performance of the proposed model imagenet models along with existing models are used. Three deep CNN models with imagenet weights, namely, Xception, Resnet 50, and Inception-V3, are used. The choice was restricted to these models due to the lower number of layers and parameters present which would provide a fair comparison for models proposed by nature-inspired algorithms used through this study. All the models were trained till the validation metrics (AUC and accuracy) would converge which in this case was approximately 20 epochs. Hence, patience criteria was kept at 2.

The following metrics were used to evaluate the proposed models.

(i) *Accuracy*: accuracy is a widely used metric for evaluating classification models. This metric is the ratio of predictions that the proposed model is correctly predicted and the total number of samples. Accuracy is defined in

$$\text{Accuracy} = \frac{\text{True Positive} + \text{True Negative}}{\text{Total Number of Samples}} \quad (10)$$

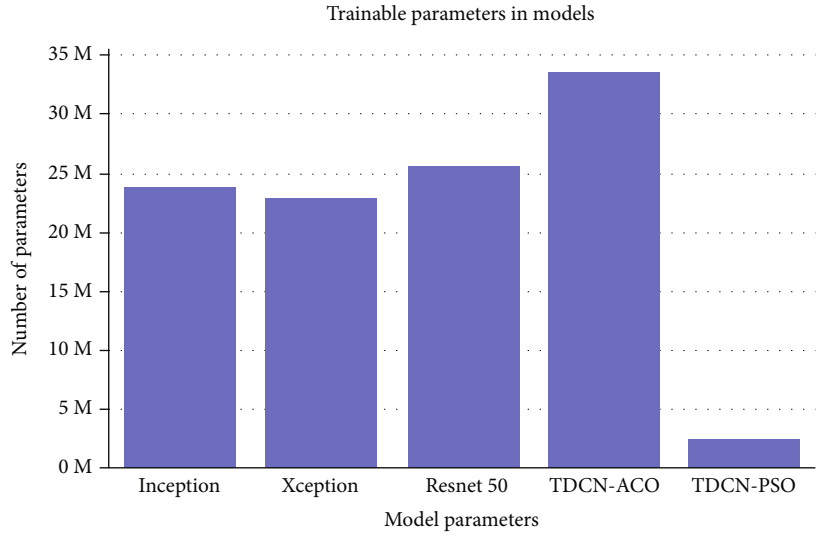

Figure 3: Size comparison of TDCN and imagenet models.

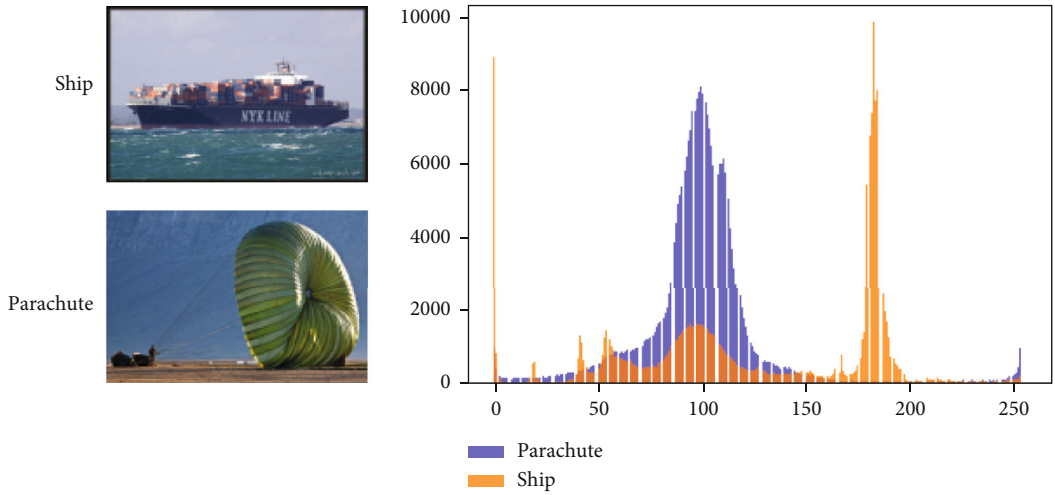

Figure 4: Pixel histogram for imagenet samples.

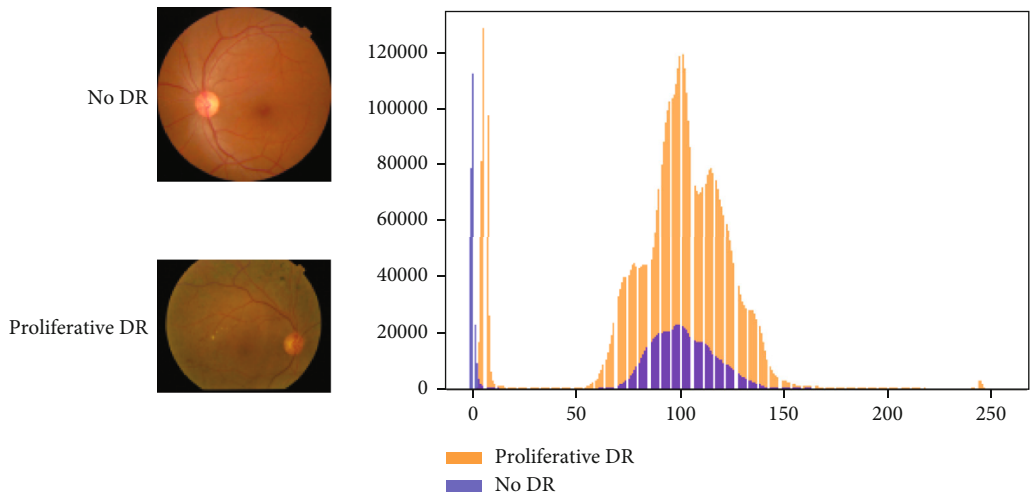

Figure 5: Pixel histogram for APTOS samples.

Table 6: TDCN-ACO results with respect to different hyperparameter settings.

| Image | Ant | Accuracy | ROC AUC | Kappa |
| --- | --- | --- | --- | --- |
| 32 | 8 | 75.7 | 0.89 | 0.710 |
| 32 | 16 | 77 | 0.905 | 0.776 |
| 64 | 8 | 76.8 | 0.888 | 0.776 |
| 64 | 16 | 78.4 | 0.907 | 0.795 |

(ii) *AUC*: AUC means the area under the curve. It is a chart that visualizes the trade-off between true positive rate (TPR) and false positive rate (FPR) for every threshold. Expression for AUC is given in

$$\text{Recall/sensitivity} = \frac{\text{True positive}}{\text{True positive} + \text{false negatives}}, \quad (11)$$

$$\text{False positive rate} = \frac{\text{False positive}}{\text{True negative} + \text{false positive}} \quad (12)$$

(iii) *Cohen's kappa score*: for the evaluation of the models, we have used a metric called Cohen's kappa coefficient which has been used for evaluating the submissions on the Kaggle competition for this dataset.

$$\text{Cohen's kappa score} = \frac{\text{pr}(o) - \text{pr}(e)}{1 - \text{pr}(e)} \quad (13)$$

Cohen's kappa is a quantitative measure of similarity between the ratings of two annotators on the same sample. It is calculated by taking into account the observed agreement and the hypothetical probability of chance agreement. Expression for Cohen's kappa is given in (13).

## 5. Results and Discussions

Table 5 represents the comparison of TDCN models with imagenet models and the other models in the literature. Figure 3 illustrates the size comparison of TDCN and imagenet models.

The results presented in Table 5 can be explained as follows:

(i) In the case of imagenet models, Xception gave the best accuracy score of 74.8%, Inception gave the best AUC ROC score of 0.91, and Resnet50 gave the best Cohen's kappa score of 0.772

(ii) TDCN-ACO provided a classification accuracy of 78.4% which is more than that observed in the case of imagenet models, AUC ROC score of 0.907, and Cohen's kappa score of 0.795. Only Inception gave better results than TDCN-ACO in terms of AUC ROC score

(iii) TDCN-PSO outperformed all the imagenet models and TDCN-ACO in terms of all the three metrics recorded. It provided a classification accuracy of 90.3%, AUC ROC score of 0.956, and Cohen's kappa score of 0.967

(iv) TDCN-PSO was significantly small in size with over 2.5 million parameters, which is about 9 times smaller than Xception and 13 times smaller than TDCN-ACO as shown in Figure 3

(v) In terms of AUC ROC, only [31] provided better performance than TDCN-PSO, while [33, 34] gave comparable results

(vi) Considering the performance over all the three metrics, TDCN-PSO gave one of the best results, while TDCN-ACO has a good place among the models and approaches compared

The imagenet dataset contains images distributed into 1000 different classes. The images from different classes differ significantly from each other in terms of features and entropy. Figure 4 shows two random images belonging to two classes, namely, ship and parachute. We plot the histogram of pixel value frequencies in order to study the distribution of features. Figure 5 shows the same histogram plot for two images from the APTOS dataset. From the two plots, it can be inferred that while the imagenet samples differ significantly in terms of distribution, the APTOS images are comparatively similar. This is one of the reasons why the tailor-fitted models TDCN-ACO and TDCN-PSO perform better than the imagenet models. Table 6 represents the TDCN-ACO results with respect to different hyperparameter settings.

Models proposed by TDCN-ACO with hyperparameters of image size 64 with 16 ants perform the best as shown in Table 6. By looking at the ROC AUC (Figure 6(a)) and confusion matrix (Figure 6(b)), it can be deduced that other than the class proliferate the model captures features that make for efficient detection. The number of parameters in the model is the highest compared to the other models but performs better than the imagenet models; therefore, one can conclude that difference in the number of parameters combats or performs better than the features captured by CIFAR training.

The AUC ROC curve and classification matrix for TDCN-PSO are shown in Figures 6(c) and 6(d). The mean training accuracy of all the runs was 89.48% with a standard deviation of 0.54%. Figure 7(a) shows that the lowest *gbest* accuracy was obtained in run 0 of 88.7% and the highest in run number 4 (the 5[th] run) of 90.31. An improvement of 0.83% accuracy over mean accuracy was obtained by running the algorithm 5 times. Figures 7(b) and 7(c) show the trend of accuracy with respect to iterations for the 5th run which achieved the best results on the validation set. For this run, a new *gbest* was found in every iteration except from iterations 4 to 6 where the *gbest* remained the same and hence the flat nature of the curve. The steepness of the curve can be associated with the amount of information being passed from the *gbest* to the new particle in the subse-quent evolutions (iterations). The best performing model (*gbest* model) out of the 5 runs provided an accuracy score of 90.3146%, AUC ROC score of 95.6, and a Cohen's kappa score of 96.67 with 2,515,406 parameters which is a substantially better score than the state-of-the-art models with a significantly smaller model.

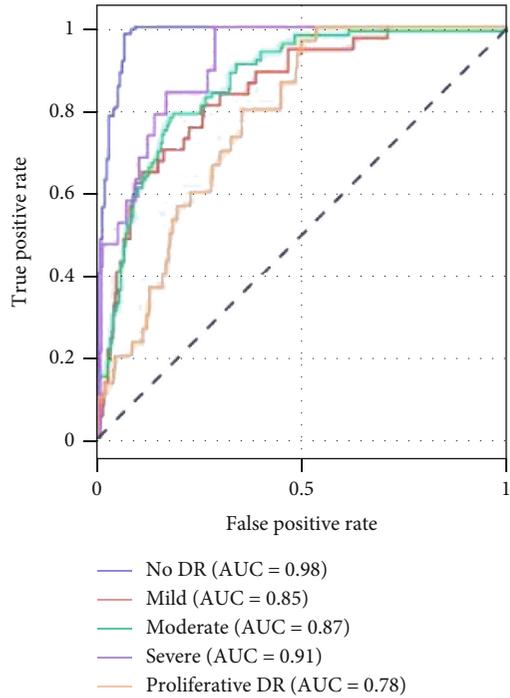

(a)

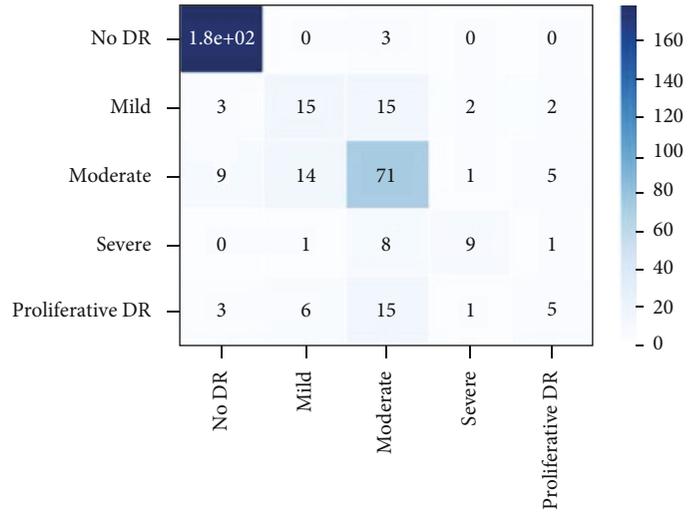

(b)

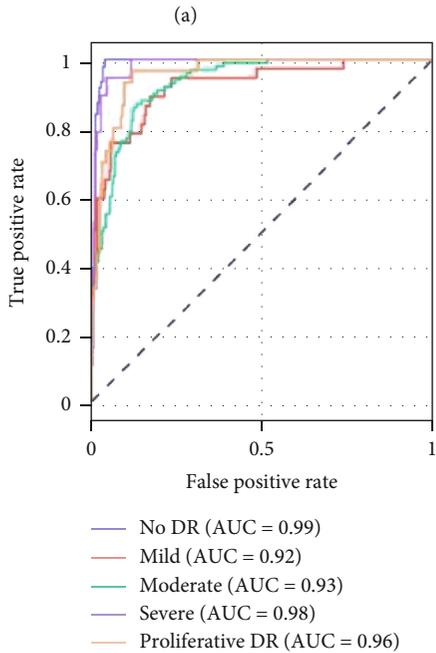

(c)

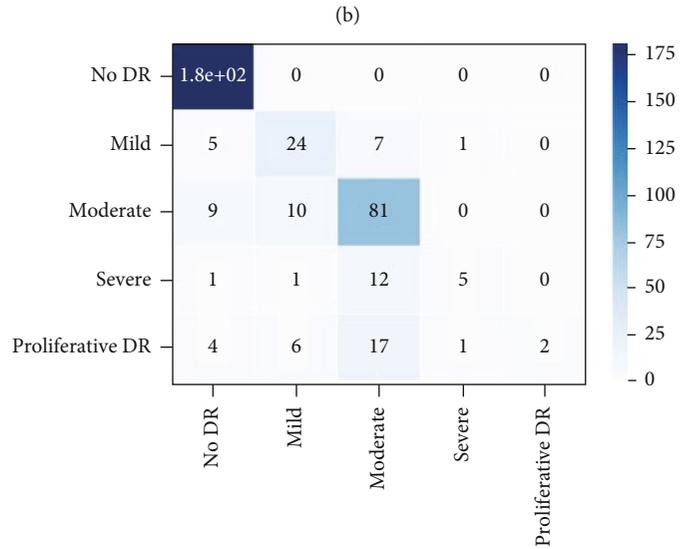

(d)

Figure 6: (a) TDCN-ACO AUC ROC curve, (b) TDCN-ACO confusion matrix, (c) TDCN-PSO AUC ROC curve, and (d) TDCN-PSO confusion matrix.

The results in Figure 6(c) show that the model found by TDCN-PSO performed really well in terms of AUC ROC

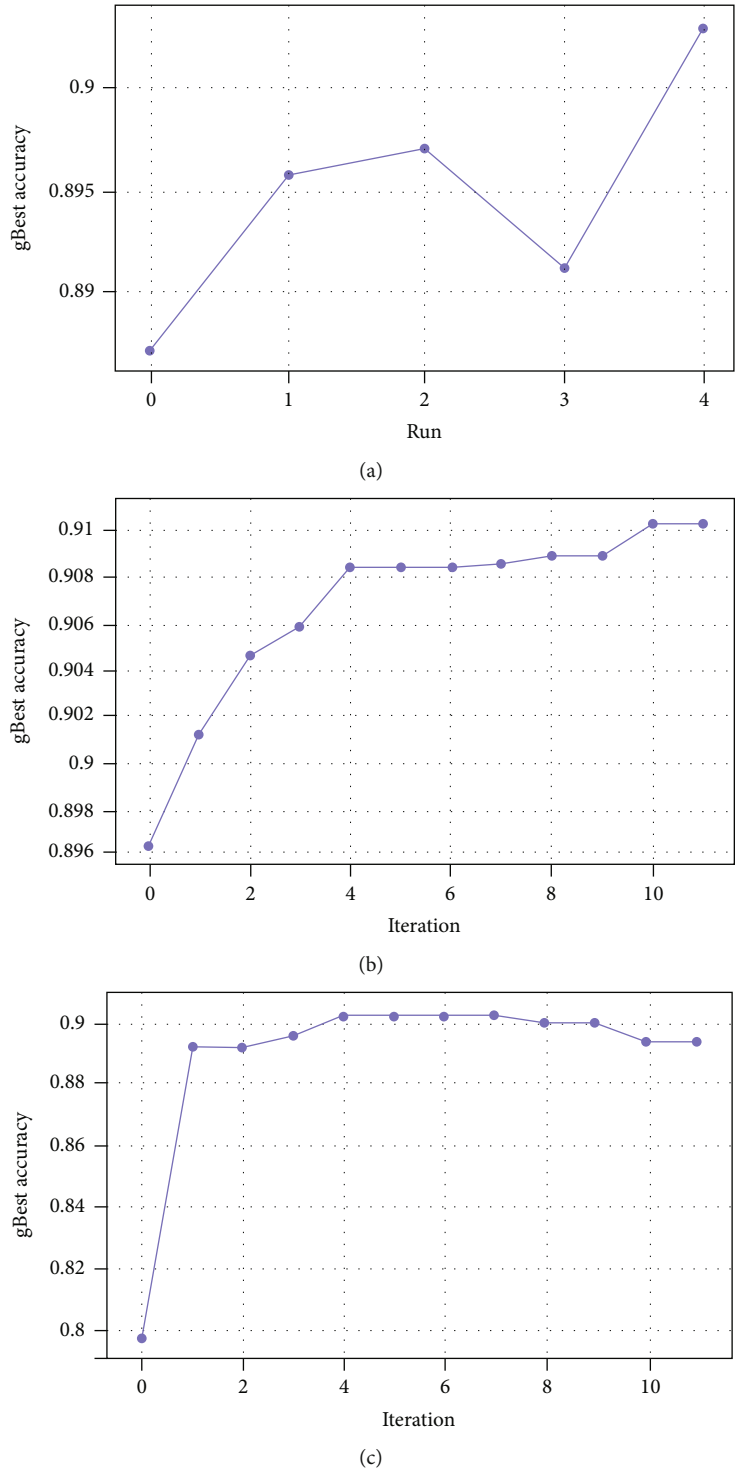

FIGURE 7: (a) Run vs. *gbest* accuracy, (b) run 5 *gbest* train accuracy, and (c) run 5 *gbest* validation accuracy.

scores. An AUC score of 0.99 for the "No DR" (no diabetic retinopathy) class conveys that the model is very accurate in classification when no diabetic retinopathy is present. In the confusion matrix obtained, it can be observed that the model was unable to produce highly accurate classifications for "severe" and "proliferative DR" classes. This was because of the skewed distribution of classes in the dataset, and the same was observed in TDCN-ACO.

## 6. Conclusions

In this paper, we have developed the lightweight deep learning-powered collective intelligence models for fundus image classification. The process of tailoring the model architecture is a problem that can be stated as a search in a multidimensional space. This paper highlights that for the chosen dataset, APTOS 2019, the models from architecture search perform better than models trained using transfer learning from imagenet weights. We utilized two swarm intelligence algorithms, namely, ant colony optimization and particle swarm optimization, to efficiently search this large space. These algorithms take a heuristic-based approach which involves deriving information from the best performing entity in the swarm. The resultant tailored deep ConvNets (TDCN) called TDCN-ACO, and TDCN-PSO outperformed the imagenet models over the metrics accuracy, AUC ROC, and Cohen-Kappa scores. TDCN-PSO while being 9 times smaller achieved an improvement of 15.5% in terms of accuracy, 0.046 in terms of AUC ROC, and 0.191 in terms of Cohen-Kappa score when compared to the best performing imagenet models in each metric. The obtained results were compared with the previous studies to show that these tailored models perform similar if not better over some metrics. The future scope of this work includes leveraging the power of other swarms existing in nature. Furthermore, tailoring an ensemble of models for datasets can be the next step to further improve the results obtained in the paper.

## Data Availability

The original contributions generated for this study are included in the article; further inquiries can be directed to the corresponding author.

## Conflicts of Interest

The authors declare that there are no conflicts of interest regarding the publication of this paper.

## Authors' Contributions

Prajjwal Gupta and Thejineaswar Guhan contributed equally to this work.

## References

[1] G. T. Reddy, M. P. K. Reddy, K. Lakshmanna et al., "Analysis of dimensionality reduction techniques on big data," *IEEE Access*, vol. 8, pp. 54776–54788, 2020.

[2] T. R. Gadekallu, D. S. Rajput, M. P. K. Reddy et al., "A novel PCA–whale optimization-based deep neural network model for classification of tomato plant diseases using GPU," *Journal of Real-Time Image Processing*, vol. 18, no. 4, pp. 1383–1396, 2021.

[3] T. R. Gadekallu, M. Alazab, R. Kaluri, P. K. R. Maddikunta, S. Bhattacharya, and K. Lakshmanna, "Hand gesture classification using a novel CNN-crow search algorithm," *Complex & Intelligent Systems*, vol. 7, no. 4, pp. 1855–1868, 2021.

[4] K. Srinivasan, L. Garg, D. Datta et al., "Performance comparison of deep cnn models for detecting driver's distraction," *CMC-Computers, Materials & Continua*, vol. 68, no. 3, pp. 4109–4124, 2021.

[5] K. Srinivasan, A. Ankur, and A. Sharma, "Super-resolution of magnetic resonance images using deep convolutional neural networks," in *2017 IEEE International Conference on Consumer Electronics - Taiwan (ICCE-TW)*, pp. 41-42, Taipei, Taiwan, 2017.

[6] U. Balakrishnan, K. Venkatachalapathy, and G. S. Marimuthu, "A hybrid PSO-DEFS based feature selection for the identification of diabetic retinopathy," *Current Diabetes Reviews*, vol. 11, no. 3, pp. 182–190, 2015.

[7] M. Bajčeta, P. Sekulić, S. Djukanović, T. Popovic, and V. Popović-Bugarin, "Retinal blood vessels segmentation using ant colony optimization," in *2016 13th Symposium on Neural Networks and Applications (NEUREL)*, pp. 1–6, Belgrade, Serbia, 2016.

[8] S. Hooshyar and R. Khayati, "Retina vessel detection using fuzzy ant colony algorithm," in *2010 Canadian Conference on Computer and Robot Vision*, pp. 239–244, Ontario, Canada, 2010.

[9] Y. S. Kanungo, B. Srinivasan, and S. Choudhary, "Detecting diabetic retinopathy using deep learning," in *2017 2nd IEEE International Conference on Recent Trends in Electronics, Information & Communication Technology (RTEICT)*, pp. 801–804, Bangalore, India, 2017.

[10] X. Zeng, H. Chen, Y. Luo, and W. Ye, "Automated diabetic retinopathy detection based on binocular Siamese-like convolutional neural network," *IEEE Access*, vol. 7, pp. 30744–30753, 2019.

[11] C. Lam, D. Yi, M. Guo, and T. Lindsey, "Automated detection of diabetic retinopathy using deep learning," *AMIA Summits on Translational Science Proceedings*, vol. 2017, pp. 147–155, 2018.

[12] A. Sakaguchi, R. Wu, and S. I. Kamata, "Fundus image classification for diabetic retinopathy using disease severity grading," in *Proceedings of the 2019 9th International Conference on Biomedical Engineering and Technology*, pp. 190–196, Tokyo, Japan, 2019.

[13] J. Bergstra, B. Komer, C. Eliasmith, D. Yamins, and D. D. Cox, "Hyperopt: a python library for model selection and hyperparameter optimization," *Computational Science & Discovery*, vol. 8, no. 1, article 014008, 2015.

[14] T. Akiba, S. Sano, T. Yanase, T. Ohta, and M. Koyama, "Optuna: a next-generation hyperparameter optimization framework," in *Proceedings of the 25th ACM SIGKDD International Conference on Knowledge Discovery & Data Mining*, pp. 2623–2631, Anchorage, AK, USA, 2019.

[15] L. Kotthoff, C. Thornton, H. H. Hoos, F. Hutter, and K. Leyton-Brown, "Auto-WEKA: automatic model selection and hyperparameter optimization in WEKA," in *Automated Machine Learning*, pp. 81–95, Springer, Cham, 2019.

[16] J. Snoek, H. Larochelle, and R. P. Adams, "Practical Bayesian optimization of machine learning algorithms," *Advances in Neural Information Processing Systems*, vol. 25, 2012.

[17] J. Wu, X. Y. Chen, H. Zhang, L. D. Xiong, H. Lei, and S. H. Deng, "Hyperparameter optimization for machine learning models based on Bayesian optimization," *Journal of Electronic Science and Technology*, vol. 17, no. 1, pp. 26–40, 2019.

[18] M. Feurer, J. Springenberg, and F. Hutter, "Initializing Bayesian hyperparameter optimization via meta-learning," *Proceedings of the AAAI Conference on Artificial Intelligence*, vol. 29, no. 1, 2015.

[19] L. Li, K. Jamieson, G. DeSalvo, A. Rostamizadeh, and A. Talwalkar, "Hyperband: a novel bandit-based approach to hyperparameter optimization," *The Journal of Machine Learning Research*, vol. 18, no. 1, pp. 6765–6816, 2017.

[20] S. Falkner, A. Klein, and F. Hutter, "BOHB: robust and efficient hyperparameter optimization at scale," *Proceedings of the 35th International Conference on Machine*, vol. 80, pp. 1437–1446, 2018.

[21] E. Hazan, A. Klivans, and Y. Yuan, "Hyperparameter optimization: A spectral approach," 2017, https://arxiv.org/abs/1706.00764.

[22] T. Hinz, N. Navarro-Guerrero, S. Magg, and S. Wermter, "Speeding up the hyperparameter optimization of deep convolutional neural networks," *International Journal of Computational Intelligence and Applications*, vol. 17, no. 2, 2018.

[23] X. Dong, J. Shen, W. Wang, L. Shao, H. Ling, and F. Porikli, "Dynamical hyperparameter optimization via deep reinforcement learning in tracking," *IEEE Transactions on Pattern Analysis and Machine Intelligence*, vol. 43, no. 5, pp. 1515–1529, 2021.

[24] J. Wu, S. Chen, and X. Liu, "Efficient hyperparameter optimization through model-based reinforcement learning," *Neurocomputing*, vol. 409, pp. 381–393, 2020.

[25] M. Rostami, K. Berahmand, E. Nasiri, and S. Forouzandeh, "Review of swarm intelligence-based feature selection methods," *Engineering Applications of Artificial Intelligence*, vol. 100, article 104210, 2020.

[26] M. Yasen and N. Al-Madi, "Improved swarm intelligence optimization using crossover and mutation for medical classification," in *2019 2nd International Conference on new Trends in Computing Sciences (ICTCS)*, pp. 1–6, Amman, Jordan, 2019.

[27] Y. Sun, B. Xue, M. Zhang, and G. G. Yen, "A particle swarm optimization-based flexible convolutional autoencoder for image classification," *IEEE Transactions on Neural Networks and Learning Systems*, vol. 30, no. 8, pp. 2295–2309, 2019.

[28] A. H. Asad, E. El Amry, A. E. Hassanien, and M. F. Tolba, "New global update mechanism of ant colony system for retinal vessel segmentation," in *13th international conference on hybrid intelligent systems (HIS 2013)*, pp. 221–227, Gammarth, Tunisia, 2013.

[29] G. Kavitha and S. Ramakrishnan, "Identification and analysis of macula in retinal images using ant colony optimization based hybrid method," in *2009 World Congress on Nature & Biologically Inspired Computing (NaBIC)*, pp. 1174–1177, Coimbatore, India, 2009.

[30] U. Bhimavarapu and G. Battineni, "Automatic microaneurysms detection for early diagnosis of diabetic retinopathy using improved discrete particle swarm optimization," *Journal of Personalized Medicine*, vol. 12, no. 2, p. 317, 2022.

[31] E. Byla and W. Pang, "Deepswarm: optimizing convolutional neural networks using swarm intelligence," in *UK Workshop on Computational Intelligence*, pp. 119–130, Springer, Cham, 2020.

[32] F. Junior, F. Erivaldo, and G. Yen, "Particle swarm optimization of deep neural networks architectures for image classification," *Swarm and Evolutionary Computation*, vol. 49, no. 62-74, pp. 62–74, 2019.

[33] M. Shaban, Z. Ogur, A. Mahmoud et al., "A convolutional neural network for the screening and staging of diabetic retinopathy," *PLoS One*, vol. 15, no. 6, article e0233514, 2020.

[34] S. H. Kassani, P. H. Kassani, R. Khazaeinezhad, M. J. Wesolowski, K. A. Schneider, and R. Deters, "Diabetic retinopathy classification using a modified Xception architecture," in *2019 IEEE International Symposium on Signal Processing and Information Technology (ISSPIT)*, pp. 1–6, Ajman, United Arab Emirates, 2019.

[35] S. Taufiqurrahman, A. Handayani, B. R. Hermanto, and T. L. E. R. Mengko, "Diabetic retinopathy classification using a hybrid and efficient MobileNetV2-SVM model," in *2020 IEEE Region 10 Conference (TENCON)*, pp. 235–240, Osaka, Japan, 2020.

[36] J. Bodapati, N. Veeranjaneyulu, S. Shareef et al., "Blended multi-modal deep ConvNet features for diabetic retinopathy severity prediction," *Electronics*, vol. 9, no. 6, p. 914, 2020.